**Determining Habitability: Which exoEarths should we search for life?**

J. Horner[1] & B. W. Jones[2]
[1] Department of Physics, Science Laboratories, University of Durham, South Road, Durham, UK, DH1 3LE
[2] Department of Physics and Astronomy, The Open University, Walton Hall, Milton Keynes, UK, MK7 6AA

## Abstract

Within the next few years, the first Earth-mass planets will be discovered around other stars. Some of those worlds will certainly lie within the classical "habitable zone" of their parent stars, and we will quickly move from knowing of no exoEarths to knowing many. For the first time, we will be in a position to carry out a detailed search for the first evidence of life beyond our Solar System. However, such observations will be hugely taxing and time consuming to perform, and it is almost certain that far more potentially habitable worlds will be known than it is possible to study. It is therefore important to catalogue and consider the various effects which make a promising planet more or less suitable for the development of life. In this work, we review the various planetary, dynamical and stellar influences that could influence the habitability of exoEarths. The various influences must be taken in concert when we attempt to decide where to focus our first detailed search for life. While there is no guarantee that any given planet will be inhabited, it is vitally important to ensure that we focus our time and effort on those planets most likely to yield a positive result.

# Introduction

Since the first planet orbiting a Sun-like star was found in 1995 orbiting 51 Pegasi (Mayor and Queloz 1995), the question of whether we will ever find life beyond our Solar System has moved firmly from the realm of science fiction to become mainstream scientific research. Research in the field of Astrobiology, spanning everything from the study of microbes to the dynamics of distant planetary orbits, has gone from strength to strength, as researchers across all fields of science come together to work on this question.

The rate at which exoplanets are discovered is rising rapidly, the numbers being continually bolstered as new techniques and telescopes come online allowing the detection of ever smaller worlds. At the time of writing, the least massive planet discovered to date around a Sun-like star, Gliese 581e, could be as little as 1.9 times the mass of the Earth. Surely, within the next few years, the first truly Earth-mass planets will be found around distant stars as projects such as Kepler (http://kepler.nasa.gov/) begin to yield their anticipated results.

If the history of astronomy tells us anything, it suggests that once one member of a population is found, many more will soon follow. For example, aside from the anomalous Pluto, the first trans-Neptunian object, the Edgeworth-Kuiper belt body (15760) 1992 $QB_1$, was discovered in 1992 (Jewitt and Luu 1993). By 2000, 374 were known, and today, early in 2010, the number has soared to 1130[1], and the discovery rate continues to accelerate. Perhaps a more telling example is the number of exoplanets known – fifteen years on from the discovery of 51 Pegasi, the catalogue of confirmed exoplanets stands at 452 planets distributed amongst 385 planetary systems[2]. Clearly, if the discovery of Earth-like planets follows this trend, we will move rapidly from knowing no exoEarths to knowing tens, or hundreds.

How, then, should we decide which exoEarths we should target in the search for life? Recent studies claiming to have detected the first molecules in the atmospheres of hot Jupiters have required significant investment in telescope time (e.g. Sing et al. 2008). To detect life-indicating molecules in the atmospheres of an exoEarth would clearly be significantly more challenging, although some of that difficulty will obviously be ameliorated by the development of new technology and the next generation of space telescopes. The same high cost of observations will hold regardless of the technique chosen to search for life, particularly since the importance of any positive discovery is such that the observers will likely want to be extra sure before making any announcement.

Regardless of what new technologies and telescopes are developed it is therefore highly unlikely that we will be able to quickly and efficiently survey all new exoEarths at once. It will therefore be vitally important to ensure that we choose the most likely candidates for the initial observations, in order to maximise our chances of finding life.

How, then, would we discriminate between different exoEarths? What determines whether one planet is more, or less, habitable than another? In this work, we attempt to summarise the key features that are currently understood to influence planetary habitability, spanning the influence of the planetary host star and its local and galactic environment in the next two sections, the various problems which can be caused by objects within the planetary system in the subsequent section, and the nature of the planet itself in the final section. Although it would be foolish to entirely prejudge where we are likely to find life, it makes sense to focus our first detailed planet-by-planet searches

---

[1] The number of known trans-Neptunian objects was determined through examining the list of those objects at http://www.cfa.harvard.edu/iau/lists/TNOs.html, accessed on 19th April 2010.
[2] The number of planets and planetary systems was taken from the Extrasolar Planets Encyclopaedia (http://exoplanet.eu/catalog-all.php) on 19th April 2010

on those that seem most likely to provide a positive detection, and so in this work we attempt to highlight potential criteria through which systems can be judged to be more or less promising for those initial observations.

## The role of the properties of single stars and stellar groups

*The variety of stars*

Stars form from fragments of interstellar clouds. As the fragment collapses it forms a circumstellar disk of gas and dust. Nearly all of the gas is hydrogen and helium. The dust, which accounts for up to a few percent of the mass of the disc, is dominated by the other 90 chemical elements. In astronomy these are called "metals", though many are non-metallic. The metallicity of the disc, or of a given star, is the proportion of these metals that it contains, relative to that contained in our Sun. Metallicity is usually expressed in terms of [Fe/H], given as the logarithm of the ratio of the iron/hydrogen ratio in the star/disk in question to that in the Sun. Therefore, stars with negative [Fe/H] are considered low metallicity, and those with positive [Fe/H] are considered high metallicity. A protostar forms in the centre of the disc, and starts with the same composition, and hence the same metallicity, as the disc.

The protostar gravitationally contracts, and as it does so it heats up, particularly in its central region. When the core reaches temperatures of order ten million K, hydrogen fusion begins, producing helium and lots of energy. This outpouring of radiation stabilises the protostar, balancing the inward pressure due to self gravity, and it becomes a star in the core hydrogen fusing phase of its life, called the main sequence phase. The star is then called a main sequence star. The Sun is a main sequence star.

Main sequence stars have a range of masses, from several tens of the solar mass down to 0.08 solar masses. Objects with lower mass are known but their interior temperatures never become hot enough for sustained hydrogen fusion to occur. These small objects are "failed stars", called brown dwarfs. It seems unlikely that such objects would house habitable exoEarths, and as such, they are not considered further in this work.

Main sequence stars are called dwarfs (see below). They are classified according to their mass. In order of decreasing mass the labels are O, B, A, F, G, K, M. The Sun is a G star. The greater the mass, the smaller the number of stars of that mass, the greater their luminosity, and the shorter the main sequence lifetime. This lifetime ends when insufficient hydrogen remains in the core to support ongoing fusion between hydrogen nuclei. As the radiation pressure resulting from that fusion is removed, the core contracts under its own gravity, and heats up. Other fusion reactions occur, not always in the core, and the star swells to become a giant at lower masses and a supergiant at larger masses. The Sun is currently 4.6 Gyr into an estimated 11 Gyr main sequence lifetime. Thereafter, it will become a giant. This is not the end, but the post main sequence evolution of the Sun will result in a series of relatively short-lived transformations which will effectively rule out life in the Solar System, and the same is thought to apply to exoplanetary systems.

You can now appreciate why main sequence stars are called dwarfs – they are significantly smaller than giants and subgiants.

*Stellar age, main sequence lifetime, and habitability*

It is certain that we will continue to find planets around stars of all ages, from those only recently formed (such as the giant planets directly imaged around the ~60 Myr old star HR 8799, Marois et al. 2008), to those well into the main sequence phase of their lifetimes. HD80606, with an age of 7.63 Gyr, is an example. On our own Earth, it is generally accepted that for the first 700 Myr or so, any life that formed would have been exterminated during the "Late Heavy Bombardment" (which we will discuss in more detail in the penultimate section, see also e.g. Gomes et al., 2005). As such, it is natural to assume that all young stellar systems are uninhabitable. However, hazards such as the Late Heavy Bombardment are likely stochastic events, which can be delayed by any length of time from the formation of a planetary system. Nevertheless, it is better to avoid stars younger than several hundred million years old. Are there any other reasons to exclude "young" stars from surveys for life?

Young stars are also well known to be significantly more active than their older counterparts. Stars on the main sequence do indeed seem to mellow with age (see e.g. Dehant et al., 2007; Guinan et al., 2005)! The younger the star, then, the greater its output of damaging high-energy electromagnetic radiation, as a fraction of its total luminosity. In addition, the amount of material shed by the star (its stellar wind) is known to be significantly greater for young stars than for old (e.g. Newkirk Jr., 1980; Wood et al. 2002, 2005). This would lead to an increased flux of charged particles into the atmosphere of a given planet, likely with potentially damaging consequences for any life attempting to develop there. On top of all this, the frequency and intensity of stellar flares is known to be far greater for young stars than old – yet another hazard to be confronted by life attempting to develop upon a given planet.

Clearly, then, the environment around very young stars is potentially highly hostile to any life (e.g. Lundin, Lammer & Ribas, 2007). Although this may not be sufficient in itself to hinder life's development, surely it is better to focus our initial attention to those stars which offer a gentler climate in which life can develop – it seems that it would be both more likely to find life there, and more likely that there would be varied and plentiful enough life to provide a strong, unambiguous signal for detection.

As a main sequence star ages, its luminosity gradually increases. Indeed, our Sun is currently thought to be some 30% more luminous than it was when it first joined the main sequence. All other things being equal, this means that the region around that star in which water could be liquid on a rocky planet's surface (the classical habitable zone, HZ) gradually moves further from the star as time passes. As such, it is perfectly feasible that a planet which is in a star's classical HZ at the current epoch would not have been in the past, and planets that were initially in that zone might now be too warm to host liquid water. We should therefore give preference to a planet that has spent hundreds of millions, or even billions, of years in the classical HZ to one that has only been habitable for tens of millions of years. It would no doubt be relatively straightforward to calculate how long a planet on a given orbit has been receiving enough energy from its parent star to host liquid water on its surface, and use this to initially focus on those worlds which have had the most time for life to develop.

This places a *lower* limit on the main sequence lifetime of the star, and we must avoid stars with main sequence lifetimes that are too short. On Earth, there is indirect evidence that life emerged about 50 Myr after the first 700 Myr of heavy bombardment, though it might have emerged a few hundred Myr later (Battistuzzi et al. 2004). If we adopt the admittedly somewhat arbitrary criterion that a star must have spent at least 1000 Myr on the main sequence to have a stable biosphere (since this would allow sufficient time for the worst excesses of stellar activity to die down, provide an opportunity for the dynamical state of the planetary system to settle down, and give time for any life on the planet to spread sufficiently to be detectable), then we must rule out the O, B, and A dwarfs, because they have main sequence lifetimes that are too short. Fortunately, with star numbers falling with increasing mass, this excludes only a small proportion of stars.

A rather larger proportion is ruled out by excluding F, G, K, and M stars that are younger that 1000 Ma. Over 10% of these stars are ruled out on this basis

*Metallicity and habitable planets*

So far, we have not paid much attention to whether suitable exoEarths could form from a circumstellar disc. Whereas the central issue here is which exoEarths are most likely to have life, it is of interest to look briefly at the requirements for an exoEarth to form in he first place. Given, as is likely, that the disk has sufficient mass to form planets, it must also have a sufficient proportion of "metals" to form exoEarths, from substances such as iron, silicates, and water. Whether it did is indicated by the metallicity of the star. "Metals" comprise all elements except hydrogen and helium, and are also known as heavy elements. These contribute about 1.6% to the mass of the Sun. It seems likely that metallicities less than about half that of the Sun might yield rocky planets no more than about 10% the mass of the Earth (i.e. no more than about the mass of Mars). Such planets are likely to lose their atmospheres to space and to the surface within about 1000 Myr, and thus probably never had the sustained capability to support life.

It is not only necessary to have heavy elements (metals), but *specific* heavy elements. For example, substantial losses of planetary atmospheres to their surface can be exacerbated by a low level of geological activity. To sustain sufficient activity to replenish the atmosphere, a large planetary mass helps, but the interior must also contain long-lived radioisotopes, notably $^{40}$K, $^{235}$U, $^{238}$U, and $^{232}$Th, to heat the interior over long periods. It seems reasonable to suppose that, all else being equal, the greater the proportion of such elements the more likely there will be a sufficient level of geological activity. Whether this is a significant further constraint is unknown.

Stars with the lowest metallicities had their origin when the Galaxy was young, before about 10 000 Myr ago. At that distant time, the interstellar medium would have had a near-primordial composition, with a low abundance of heavy elements. The stars forming from this medium were comparably depleted. A proportion of G, K, and M dwarfs must be at least as old as 10 000 Myr – 20% is one estimate – and this proportion is therefore unlikely to have exoEarths. Subsequently, the short-lived massive stars, in which thermonuclear fusion had increased the proportion of heavy elements, enriched the interstellar medium as they lost mass in their giant, supergiant and subsequent phases. Younger stars will therefore have been born in clouds enriched in heavy elements, and are thus more likely to have habitable planets, though some of these stars will be too young for life to have emerged in their planetary systems.

Stars not yet ruled out from having planets on which life might be present are thus the higher metallicity main sequence stars of spectral types F, G, K, and M (i.e. with masses less than about twice that of the Sun), and main sequence lifetimes exceeding about 1000 Myr, but older than about 1000 Myr.

*Stellar variability*

It is not just young stars that can be variable. To some extent, all stars are likely variable – our Sun, for example, varied in luminosity by of order 0.05% in the period 1978 – 2006 (spanning just over one complete 22-year dual-peaked solar cycle), albeit with short term variations due to spots and faculae spanning a range a factor of 10 larger in the period since 1978 (see e.g. Fig 1, Foukal et al., 2006). Over longer timescales, it seems reasonable to think that the true variability of our Sun is somewhat higher. For example, Lean (2000) examines the variation in solar insolation since the Maunder minimum, suggesting an overall increase of ~0.2% in that time. Such long-term variability has been used by some authors to study the relationship between climate change and the activity of the Sun (e.g. Lean, Beer & Bradley, 1995; Sofia & Li, 2001; Rind, D., 2002; Feulner & Rahmstorf, 2010). Such work shows how even small variations in the luminosity of a star can have measurable effects on the climate of planets orbiting around it. Some stars, though, are far more variable than others. Examples include the famous Cepheid variable stars, old stars which have recently moved from the main sequence, whose atmospheres expand and contract periodically giving rise to luminosity variations that may span a factor of two or more. The period of these variations is strongly linked to the luminosity of the star in question, allowing Cepheids to be used as exceptional "standard candles", providing a yardstick to measure distances within the Universe. Some other stars vary by a far greater amount. For example, over a period of 332 days, the luminosity of the star Mira (o Ceti) varies by up to a factor of 4,000, and this is far from atypical. Other kinds of variability also exist for stars – some, for example, are prone to enormous stellar flares (a good example being our nearest stellar neighbour, Proxima Centauri).

Clearly, stars that are hugely variable in luminosity, or are prone to enormous stellar flares, represent bad targets for the search for life. Indeed, it is unlikely that Earth-like planets will be discovered around such stars in the near future, even if they do exist, since the intrinsic variations in the star itself mask any planetary signal. Two obvious questions therefore are - how variable must a star be in order to make an otherwise habitable planet in orbit around it inhospitable, and what degree of variability is enough to lower the habitability sufficiently to make a planet a poor first choice for observation? We should certainly focus our initial search for life on planets around stars that observations or modelling indicate are particularly stable in their output (see, for example, Eddy et al. 1984, and the Kepler website at http://www.kepler.arc.nasa.gov/). Again, the strategy is to maximise the chances for a rapid/straightforward detection of life.

*ExoEarths in the classical HZ of M dwarfs*

The abundant, long-lived M dwarfs increase considerably the number of places where we might find exoEarths in the classical HZ. Some problems have been raised, but none is fatal. We discuss two erstwhile problems.

First, the low luminosity of M dwarfs results in the classical HZ being close to the star. For the most luminous M dwarfs it still only stretches from about 0.2 to 0.4 AU. An exoEarth at such a distance would have had its rotation slowed by the gravitational gradient across it (tidal forces) until a stable configuration is reached in which it keeps the same face towards the M dwarf. Such a configuration is known as a spin-orbit resonance, with the orbiting body completing an integer number of spins on its axis in the same period taken to perform an integer number of orbits around the parent. The Moon keeps the same face towards the Earth because of the tidal force exerted on the Moon by our planet, and is said to be trapped in a 1:1 spin-orbit resonance (one spin in the same period as one orbit). An exoEarth keeping the same face towards the M dwarf could have such a cold dark side that the whole atmosphere could freeze into this cold trap. Water could freeze on the dark side and evaporate from the star-facing side. However, this inimical-to-life atmospheric and water freeze-out would not happen if a substantial atmosphere were present. For example, an atmosphere with a surface pressure about a tenth of that on the Earth would prevent freeze-out provided that it consisted largely of the greenhouse gas $CO_2$ (Heath et al. 1999). At somewhat higher pressures liquid water would be present over at least part of the surface, perhaps beneath a thin crust of ice. Thus, tidal slowing does not necessarily prevent the formation of a surface biosphere.

It should be noted that it is possible for objects to be trapped in spin-orbit resonances other than 1:1. Although the 1:1 is most common (the effect of that particular resonance being the strongest), other resonances may be enough to arrest the tidal spin-down of planetary bodies. Within our own Solar System, the planet Mercury is trapped in a 3:2 spin-orbit resonance with the Sun (e.g. Correia & Laskar, 2004). However, capture into such higher-order resonances is typically highly unlikely unless the orbital eccentricity of the secondary is sufficiently high to result in a significant difference in tidal effects between perihelion and aphelion. In the case of Mercury, Correia & Laskar invoke the chaotic excitation of the planet's orbit to an eccentricity greater than 0.325 in order to facilitate efficient capture to this spin-orbit resonance. Planets trapped in such higher-order spin-orbit resonances would experience a slow diurnal cycle, and so should not necessarily be ruled out in the search for habitable worlds. That said, capture to such resonances seems sufficiently unlikely for planet's on near-circular orbits (such that insolation does not vary prohibitively over the course of one planetary year) that we do not consider them further.

Second, M stars are more variable in luminosity than the more massive main sequence stars. There are two mechanisms. First, flares, lasting typically the order of a minute, increase the luminosity, including the biologically damaging UV and X-ray wavelengths which can increase by a factor of order 100 (e.g. Scalo et al., 2007). However, even during a strong flare, the X-ray and UV flux from an M dwarf remains feeble, and poses no threat to a surface biosphere. Second, all stars have transient, cool patches on their surface, starspots. In the case of M dwarfs these are comparatively large and can cause a few tens of percent decrease in luminosity lasting up to a few months. However, even a modest atmosphere on an exoEarth would prevent such a decrease from doing much harm to a surface biosphere (e.g. Heath et al. 1999).

It should be noted that the slow rotation of a planet keeping one face towards an M dwarf could mitigate against it having a strong magnetic field (Russell et al. 1979). In such a scenario, the energetic particles in the wind that all stars emit would not be deflected, and would thus impact the upper atmosphere at all latitudes. These high speed particles collide with molecules in the planet's upper atmosphere. Among the collision products are X- and gamma rays that reach the surface where they would increase the mutation rate in any biosphere, though as a biosphere would have evolved in such an environment the effect could be positive, by promoting evolution. However, we note that the discussion of the effect of such slow rotation on planetary magnetic fields is still open. Stevenson (2003) explicitly states that "slow rotation may be more favourable for a dynamo than fast rotation", since it can lead to an increase in convective velocity within the planet's outer core. The caveat is that the Coriolis force must remain dynamically important – if the planet rotates too slowly for the force to play a significant role on the movement of the outer core, then presumably this would significantly lessen the likelihood of that planet developing a strong dynamo-driven magnetic field. (For an introduction to the Coriolis force see Wallace & Hobbs (2006) a.). The potential importance of a planetary magnetic field in constraining that planet's habitability is discussed in more depth later in this work.

*Stellar companions*

Thus far we have considered single stars. But somewhat more than half of the stars like the Sun are accompanied by a second star; these constitute binary stars. In a few cases the star has more than one stellar companion; these constitute multiple star systems. From now on we use the term "multiple star systems" to include binary stars. (Note that by Sun-like we mean F, G and K dwarfs, in order of descending mass. The Sun is a G dwarf. O, B, and A dwarfs are more massive than F dwarfs, but are rare. M dwarfs are less massive than K dwarfs, and though the most abundant stars in our galaxy, seem almost entirely present as isolated stars. However, it is possible that this is a selection effect, with small/faint stellar companions being hard to detect, particularly at wide orbital separations).

Multiple star systems can have planets. Among the 400 or so known exoplanetary systems roughly a quarter are in such systems. However, this is a lower limit, due to observational selection effects, which make close binaries difficult to investigate by the fruitful radial velocity technique. Also, faint stellar companions, such as M dwarfs or white dwarfs, are difficult to detect. Let us therefore concentrate on stars within 20 parsecs (65 light-years) of the Sun, where the census is likely to be complete. There are 38 exoplanetary systems within this range, of which 11 are in double star systems, and 1 in a triple star system. These 12 are detailed in Table 1 (Desidera and Barbieri 2007). Note that 12/38 is 32%, but the sample is small, and with a larger complete sample this proportion is expected to rise, because models of planetary formation in multiple star systems generally indicate weak constraints imposed by the other star(s) – see below.

Table 1  Exoplanetary systems in multiple stellar systems within 20 parsecs (65 light-years)

| Stars (The separation of the stars exceeding a few tens of AU is, in most cases, the projected distance on the sky.) | Distance/ parsecs | Known planets min mass /$M_J$ (1) |
|---|---|---|
| 54 Piscium, K dwarf with one planet, with a brown dwarf at  476 AU (2) | 11.1 | 0.227 @ 0.30 AU |
| 55 Cancri, G dwarf with 5 planets, with an M dwarf at 1065 AU | 12.6 | 0.034 @ 0.038 AU |
| | | 0.824 @ 0.115 AU |
| | | 0.169 @  0,24 AU |
| | | 0.144 @ 0.781 AU |
| | | 3.835 @ 5.77 AU |
| Upsilon Andromedae, F dwarf with 3 planets, with an  M dwarf at 750 AU | 13.5 | 0.69 @ 0.059 AU |
| | | 1.98 @ 0.83 AU |
| | | 3.95 @ 2.51 AU |
| Gamma Cephei, orange subgiant with 1 planet, with an M dwarf at 19.4 AU (3) | 13.8 | 1.60 @ 2.04 AU |
| Tau Boötis, F dwarf with 1 planet, with an M dwarf at 240 AU | 15.6 | 3.9 @  0.046 AU |
| GJ 3021, G dwarf with 1 planet, with an M dwarf at 68 AU | 17.5 | 3.32 @ 0.49 AU |
| HD 189733, K dwarf with 1 planet, with an M dwarf at 216 AU | 19.5 | 1.15 @ 0.0312 AU |
| Gliese 86, K dwarf with 1 planet, with a white dwarf at 18.4 AU (4) | 10.8 | 4.01 @ 0.11 AU |
| HD 147513, G dwarf with 1 planet, with a white dwarf at 4450 AU | 13.0 | 1.0 @ 1.26 AU |
| Epsilon Reticuli, subgiant with 1 planet, with a white dwarf at 250 AU (3) | 18.2 | 1.28 @ 1.18 AU |
| 83 Leonis, two G dwarfs, projected separation 515 AU; 1 planet (5) | 17.7 | 0.109 @ 0.123 AU |
| Gliese 777, G dwarf with 2 planets, with a pair of M dwarfs at about 3000 AU | 15.9 | 0.057 @ 0.128 AU |
| | | 1.50 @ 3.92 AU |

Notes
1 Most have been observed only through radial velocity measurements, which yields minimum masses. *On average*, the actual mass is about 1.3 times the minimum.
2  A brown dwarf is more massive than planets, but less massive than stars.
3 A subgiant is a star that has recently left the main sequence, in this case it was a G or K main sequence star.
4 A white dwarf is the hot compact remnant of a F/G/K main sequence star at the end of its life.
5 The planet orbits the less massive of the two stars.

You can see from Table 1 that exoplanets are found in a variety of multiple star systems. This variety also applies to the larger but incomplete sample provided by all of the 400 or so known exoplanetary systems. The planets themselves are predominantly giants with very few approaching the mass of the Earth. This is because the less massive the planet the smaller its effect on the motion of its star. The important point is that if giant planets can form with as much facility in multiple planet systems as in the case of isolated stars, then it is very likely that the same applies to exoEarths in the classical HZ of a star. Whether the exoEarth survives ejection from the classical HZ depends on the giant(s) in the system.

What do orbital simulations tell us? David et al. (2003) considered the orbital stability of an exoEarth 1 AU from a solar-type star, which is in the classical HZ of the star. They found that, if the Earth could form, then it would have an orbit stable for at least 4.6 Gyr (the present age of the Solar System), provided that
- if the orbital eccentricity of the companion star is close to zero, then it would need to have an orbital semi-major axis of at least 2.5 AU if its mass was 0.001 times that of the Sun, ranging up to at least 6 AU if its mass was half that of the Sun
- if the orbital eccentricity of the companion star is 0.9-0.95 then it would need to have a semi-major axis greater than 50 AU over the mass range 0.001- 0.5 times the mass of the Sun.

They describe scaling laws to apply these results to systems with different masses of the primary star.

They do not consider stellar separations less than 1 AU, but note that two stars separated by a small fraction of 1 AU, very close binaries, would allow long-term orbital stability of an exoEarth in the classical HZ (Holman and Weigert 1999).

Note that if the companion star is in an orbit highly inclined to the orbital plane of the planetary system, then even at a separation of hundreds of AU it can destabilize the planetary system. No systems in the process of disruption have been seen.

Overall, it is re-assuring that exoEarths can have stable orbits in a wide variety of binary star systems. But could exoEarths form in such a wide variety of systems?

*Formation of planets in binary star systems*

Desidera and Barbieri (2007) have concluded that if the two stars in a binary system approach each other no closer than about 200 AU then the circumstellar discs of dust and gas form planets as readily as do the discs around isolated stars. At closer separations the circumstellar discs are truncated, though planets still form, albeit experiencing an increase in orbital eccentricity due to the gravity of the companion star. Modelling has shown that, for a solar mass star, giant planets can form even when the (presumably lower mass) companion approaches to about 50 AU (Pfahl and Muterspaugh 2006). At yet closer separations, numerical simulations by Quintana et al. (2007) indicate that binaries containing G, K, or M stars with separations down to about 10 AU could have circumstellar discs extending out to at least 2 AU from which could form a few planets with masses up to about that of the Earth. *Giant* planets would not form, because the disk does not extend beyond the ice-line, beyond which the abundant water would condense to provide the massive cores that acquire disk hydrogen and helium to form a giant planet. (Note that in isolated stars and in multiple star systems, the occurrence of giant planets closer to the star than the ice-line is doubtless due to inward migration, through gravitational interactions between the (growing) giant and the remains of the circumstellar gas and dust disc.)

Close binaries, with minimum separations of a few AU, can have a circumstellar disk encompassing both stars. Simulations by Quintana and Lissauer (2006) show that if the two stars are in a low eccentricity orbit with a semi-major axis no greater than about 0.2 AU then a planetary system broadly resembling the Solar System could result. Slightly larger separations and/or eccentricities result in the retention of no planets. Note that close binaries are generally excluded from searches for planets by the radial velocity (RV) technique, which has discovered the great majority of the exoplanetary systems. In this technique, the presence of one or more planets is inferred from cyclic variations in the speed of the stars along our line of sight, which results from the gravity of the orbiting planet(s). The large speed in orbit of the two stars produces its own cyclic variations in the radial speed of the stars, thus obscuring the planet-produced variation. So far, no planets have been discovered orbiting close binaries. Note that exoEarths induce far smaller RV variations than giant planets, which is why no exoEarths have been discovered by the RV technique (nor, indeed, by any other technique).

*The search for exoplanetary systems in star clusters*

As well as multiple star systems, stars in our Galaxy are often found in large groups. There are open clusters, consisting of a few hundred stars, and globular clusters, with the order of a million stars (Figure 1).

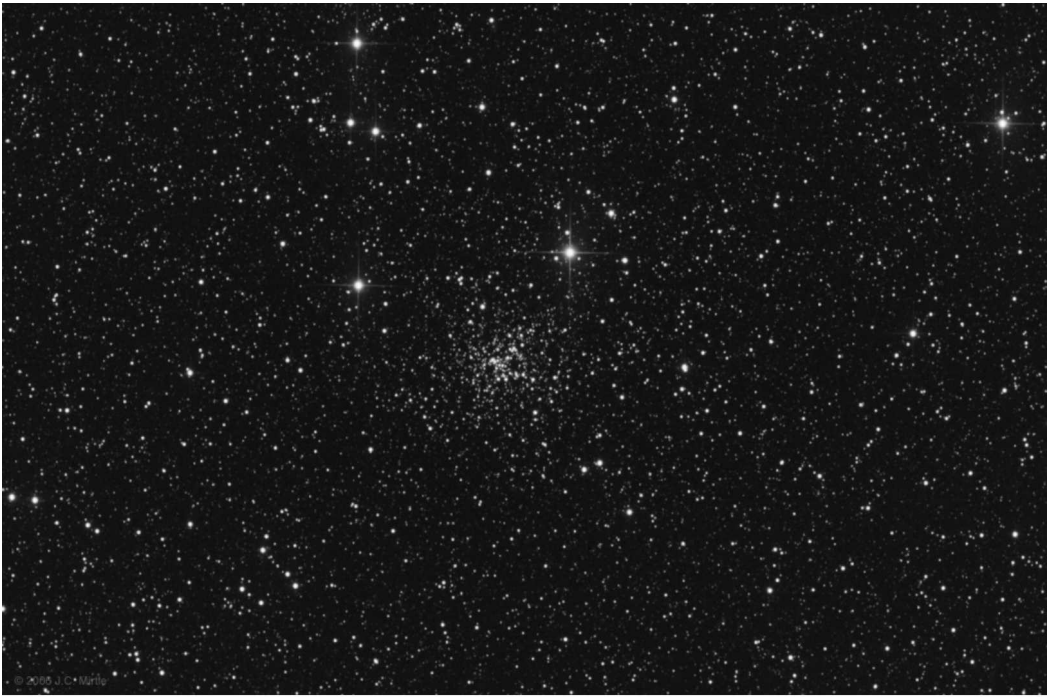

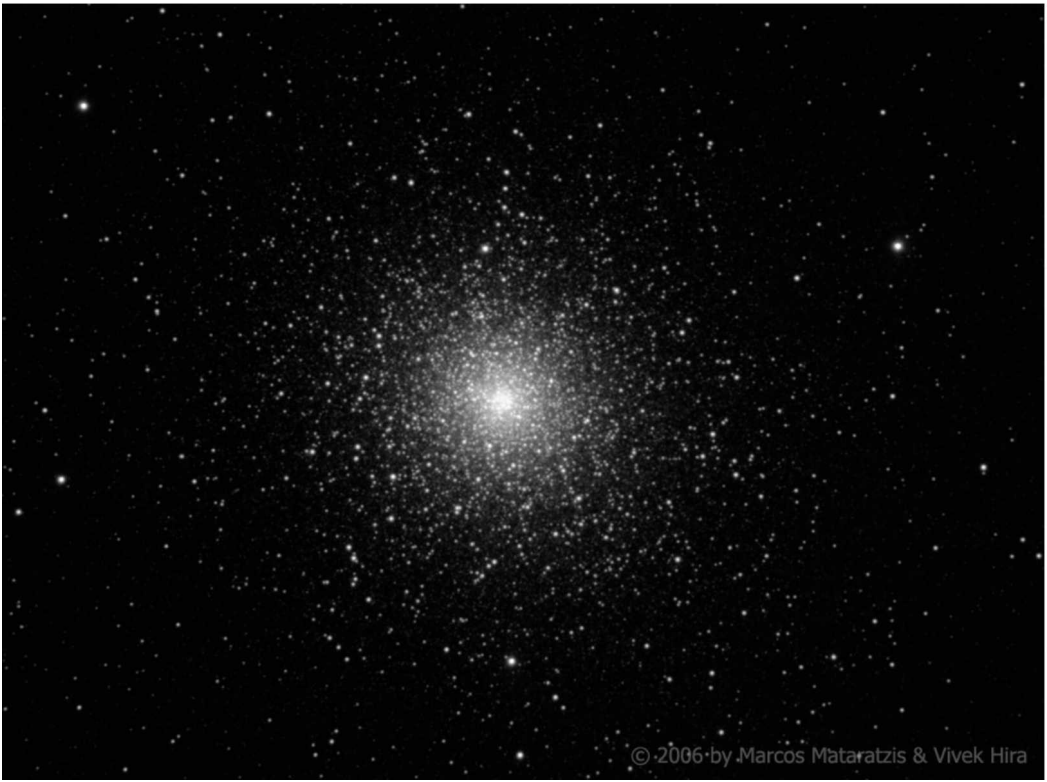

Figure 1 Top: the open cluster NGC6819 that consists of about 150 stars, each about 2400 Myr old. It is nearly 8000 light years away and is about 12 light years across. (John Mirtle, Calgary, Alberta, Canada). Bottom: The globular cluster 47 Tucanae that consists of about a million ancient stars. It is about 16 700 light years away and about 120 light years across. (Marcos Mataratzis and Vivek Hira)

Whereas the 200 or so globular clusters known in our galaxy are all ancient, having formed about thirteen billion years ago when the Galaxy was very young, open clusters are much younger, and are still forming. Whereas globular clusters are stable, open clusters gradually disperse on a time scale of order a few million years, though some are older (Figure 1). Globular clusters contain little interstellar matter and exhibit no star formation. Star formation over the past several billenia has been occurring in open clusters, which form from comparatively dense fragments of interstellar gas and dust. It is thought that the Sun was born in an open cluster 4600 Myr ago, long since dispersed.

The open cluster in Figure 1 has been scrutinised for planets around its main sequence stars by Street et al. (2003), using the transit method, in which periodic dips in the apparent brightness of a star are produced by a planet in an orbit presented edgewise to us when it passes between us and its star. However, even though NGC6819 consists of high metallicity stars, the survey, though detecting the transits of brown dwarfs, failed to reveal any transits of Jupiter size planets, in spite of the expectation of a few based on statistics from the solar neighbourhood. This might indicate that the solar neighbourhood is particularly well endowed, or be the result of the RV method, that has yielded most of the local discoveries, being less able to detect radial velocity variations at greater range.

A transit survey by Weldrake et al. (2005) of 21 920 main sequence stars in the outer regions of the globular cluster 47 Tucanae (Figure 1), also found none, in spite of a prediction of seven. Earlier, in 2000, the Hubble Space Telescope was used to search for transits of 34 091 main sequence stars in the dense core, again without success, even though 15-20 were predicted. In the latter case this absence is presumably because these ancient stars have very low metallicities, and the comparatively high spatial density of stars in such clusters might prevent planets forming, or result in close encounters between stars which would eject any planets. In the former case, the lower spatial density of stars indicates that the lower metallicity is the reason. However, a planet of a few Jupiter masses discovered in a binary system consisting of a pulsar and a white dwarf in the globular cluster M4 indicates that in rare circumstances globular clusters might have a few planets.

To date, there have been over 20 transit searches, covering open and globular clusters, all without much success, though in 2007 indirect evidence was obtained for planets forming around a star in the Pleiades open cluster. There is no good explanation for the dearth of discoveries (Janes and Kim 2008). The subject of exoplanets is very young, and there are many uncertainties.

## The Galactic Habitable Zone

Just as a star has a habitable zone, so does our galaxy, the Galaxy. Figure 2 is a sketch of the Galaxy showing an edgewise view of the major structural components – the thin and thick discs, the central bulge, and the halo. Note that the thick disk permeates the thin disc, and is recognisable there as a distinct population of stars. The numbers of stars that constitute the thin disc, thick disc, nuclear bulge, and halo are in the approximate ratios 100:20:10:1, and so the thin disk accounts for about three-quarters of the stars in the Galaxy. The Galactic habitable zone is defined in terms of the likelihood that habitable planets could be present in each structural component.

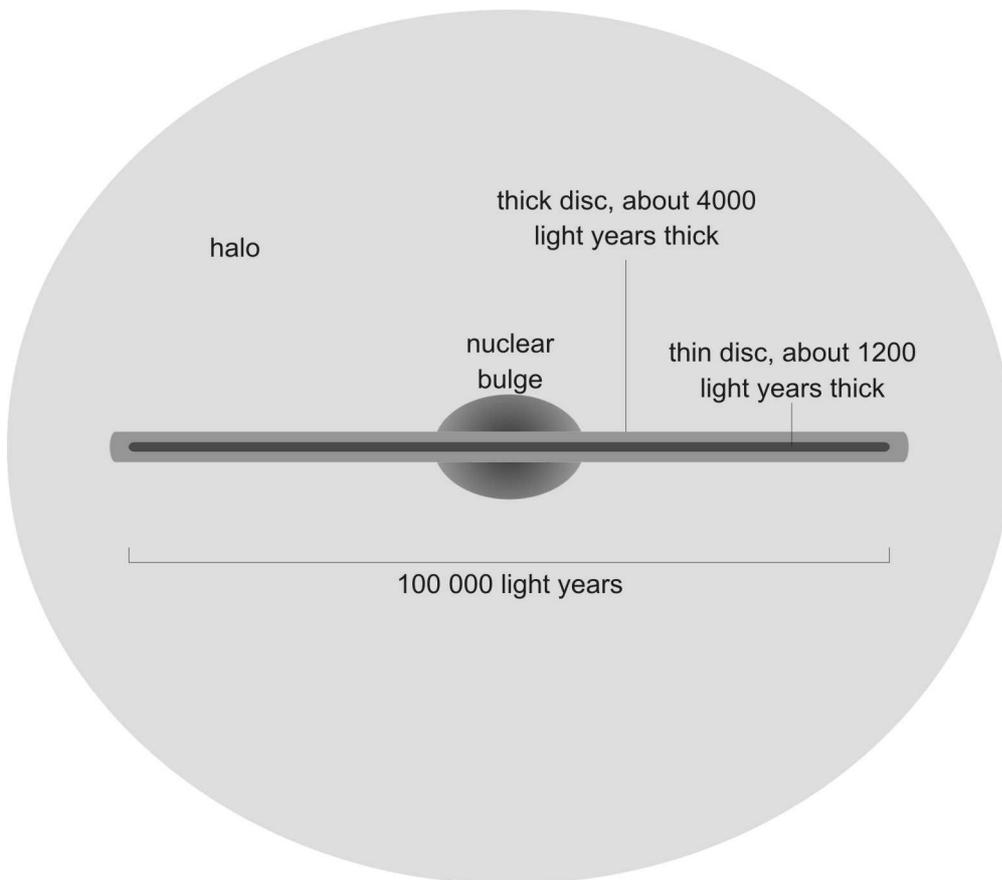

Figure 2    A sketch of the Galaxy, viewed edgewise, showing the main structural components. The boundaries are not as sharp-edged as shown here.

The metallicity of the medium from which a star and its planetary system formed is of prime importance, as was outlined earlier, with metallicities less than about half that of the Sun perhaps unlikely to yield suitable planets. The thin disk has a long and continuous history of star formation – it is where the young open clusters are found – and therefore the metallicity of its interstellar medium was raised early in Galactic history and has continued to rise since. It is the prime location for stars with habitable planets. In its outer regions it is less enriched, so suitable planets might be scarcer there. The thick disk has a much higher proportion of old, low metallicity stars, and so habitable planets are probably rarer. The halo is dominated by stars even older than those in the thick disc, with very low metallicities, and, as noted above, habitable planets are thus likely to be rare. About 1% of the halo stars are in globular clusters (Figure 1), which are found also in the nuclear bulge. In the nuclear bulge star formation peaked some time ago, but is still continuing. Habitable planets might be common, though the metallicity is somewhat lower than in the thin disc, with uncertain consequences. See Frogel (1988) and Mezger (1996) for extensive reviews of the nuclear bulge, and Wyse et al. (1997) for a review of galactic (nuclear) bulges in general.

As well as metallicity, there are two other factors that affect planetary habitability, transient radiation events and gravitational disturbance. A proportion of planetary systems will have been sterilised by transient radiation events, such as supernovae, and a further small proportion will have been disrupted by the gravitational disturbance of nearby stars. Transient radiation events occur throughout the disc, but in the outer disk are well separated and rare. They are more pervasive in the nuclear bulge and the inner disk and probably reduce significantly the habitability there. They must also have reduced the habitability of globular clusters, where massive stars long ago died in supernovae and bathed the cluster with lethal radiation. Gravitational disturbances are also significant in the bulge and (as noted above) in the inner regions of globular clusters, because they are comparatively densely packed with stars.

The thin disk is where the greatest proportion of stars is likely to have habitable planets, particularly in an annulus that excludes its outermost and innermost regions. The Sun is in this annulus! With the thin disk accounting for about three-quarters of the stars in the Galaxy we thus have to exclude somewhat more than a quarter. Of the remainder, some proportion is unlikely to have planets on which life developed, for the various reasons given earlier.

As an *upper limit*, roughly half of the stars in the Galaxy could have habitable planets if M dwarfs are included, otherwise the proportion is 5-10%. It must be emphasised that these are *very* rough figures.

## Dynamical Effects and Debris

*Orbital parameters and stability*

Many of the exoplanets discovered to date move on orbits far different to those occupied by the planets in our Solar System. Many of the discoveries have brought new surprises, and our understanding of how planetary systems form and evolve has changed dramatically as a result. With such a wide range of possible scenarios, it is vital that the orbital stability and evolution of an exoEarth be examined in some detail before selecting it as a prime site to look for life.

Something which can quickly be determined observationally for a given planet is the eccentricity of its orbit. The more eccentric the orbit, the greater the difference in insolation between periastron and apastron. In addition, planets move faster at periastron than apastron, with the variation in speed described by Kepler's second law (the radius vector, a line joining the centre of the planet and the centre of the star, sweeps out equal areas in equal times). The more eccentric the planetary orbit, then, the greater the difference between orbital speed at periastron and apastron. So, as an extreme case, a planet with a semi-major axis that places it smack in the "habitable zone", but with a highly eccentric orbit, will move so that it spends a very short period of time near periastron, receiving a huge amount of radiation from the star, and a very long period out near apastron, with a greatly reduced flux. In other words, the annual variation in surface temperature could be so dramatic as to render the planet significantly less habitable than would otherwise be the case.

The obvious conclusion, therefore, is that we would want to look, initially, at planets located on low eccentricity orbits. However, just because a planet is currently moving on a circular orbit does not mean that it always has done, or will for the foreseeable future. The temporal evolution of a planet's orbit is affected by every other object in the planetary system, but is dominated by the influence of the most massive, which could be other planets, and any companions the star has.

A good example of a recently discovered planetary system believed to be dynamically unstable is the three planets imaged around the youthful star HR8799. Although the star is some ~60 Myr old (Marois et al., 2008), studies of the three giant planets in orbit around it have suggested that their orbits are unstable on timescales of hundreds of thousands of years, or even less (Fabrycky & Murray-Clay, 2010), although recent work (Marshall, Horner & Carter, 2010, elsewhere in this proceedings) suggests that some plausible configurations for the system allow stability on at least Myr timescales.

When we consider the habitability of an exoEarth, we must therefore examine the long term dynamical stability of its orbit. Just because the orbit appears suitable now does not mean that it was so in the past, or will remain so in the future: any decision on which exoEarth to study must take into account the long term dynamical variation of the planet in question.

*Climate stability*

Even if the orbit of the planet is stable on a macroscopic scale, protected from any great excursions in semi-major axis or eccentricity, the perturbations of the other planets in the system could play a significant role in determining its habitability. On Earth, subtle periodic variations in the eccentricity and inclination of our orbit, coupled with small variations in the tilt of our rotation axis, have been shown to be intimately linked to the recent period of repeated glaciations and inter-glacial periods. These variations, known as the Milanković cycles, show that even small scale perturbations could play an important role in determining the degree to which a planet is habitable. Were our Solar System laid out slightly differently, it is quite plausible that these variations would be significantly larger, or happen over a shorter timescale, both of which could significantly alter the habitability of our planet.

It just so happens that our Earth only experiences fairly small variations that occur over a fairly lengthy period of time. Waltham (2010) used Monte Carlo simulations to show that up to 98.5% of randomly generated planetary systems that host an Earth-analogue planet would experience significantly more rapid and potentially more extreme Milanković cycles than we do, suggesting that, statistically, our Earth might be unusually favourable for the development and survival of life.

Fortunately, calculations to determine the periodicity and severity of such variations for a given system are not particularly computationally intensive, and so long as we have a reasonable knowledge of the make-up of an exoEarth's host system, we should be able to rapidly determine how severe and rapid and climate change resulting from such effects would be, and draw some quick conclusions about the degree to which the planet is optimally habitable.

*Planetary Shielding*

It has long been believed that giant planets, such as Jupiter, can act to shield potentially habitable planets from potentially hazardous objects, thus proving a great boon to the development of life. Until recently, however, this idea had received little serious study. A series of recent papers (Horner & Jones 2008a, 2008b, 2009; Horner, Jones & Chambers 2010) has begun the detailed study of this problem. In our own Solar System, at the current epoch, there are three groups of potentially hazardous objects.

The Near Earth Asteroids (Chapman 1994; Bottke et al., 2002; Morbidelli et al. 2002) are thought to make up ~75% of the current impact hazard at Earth. These objects are primarily sourced from the Asteroid belt, and the great bulk are collisional fragments of larger parent bodies. Once such a fragment is created in the main belt, its orbit evolves in response to both non-gravitational forces (such as the Yarkovsky and YORP effects (e.g. Morbidelli & Vokrouhlický, 2003; Vokrouhlický & Čapek, 2002)) and the gravitational perturbations of the planets, particularly Jupiter. Most such material eventually strays into one of the many secular or mean-motion resonances which are spread through the belt, at which point they are driven inward to threaten the terrestrial planets. Clearly, then, changing the precise architecture of the outer Solar System (the locations, orbits, and masses of the giant planets) would act to significantly alter the efficiency with which such objects are flung inwards towards the Earth. As such, Horner & Jones (2008) examined the effect of varying the mass of Jupiter on the impact flux the Earth would experience from asteroidal material. Far from simply being a shield, they found that the relationship between the mass of Jupiter and the impact flux at Earth was reasonably complex. For very small "Jupiter" masses, the impact flux was relatively low, rising to a huge peak when "Jupiter" was around the mass of Saturn, then declining so that at the mass of Jupiter it was still high. The impact flux continued to decline as the mass of "Jupiter" increased.

The Jupiter-family comets (Levison & Duncan, 1997) make up a further significant contribution to the impact hazard for terrestrial planets. As their name suggests, their orbits are dominated by the influence of the planet Jupiter, with their aphelia located in the vicinity of the planet's orbit. The proximate source of the Jupiter-family comets are the Centaurs (e.g. Horner et al. 2003, 2004a, b; Tiscareno & Malhotra 2003; di Sisto & Brunini 2007), objects moving on dynamically unstable orbits with perihelia between the orbits of Jupiter and Neptune. These objects themselves display significant dynamical instability, and so must be replenished from at least one source reservoir somewhere in the outer Solar System. Over the past fifteen years, a number of regions have been proposed to be the main source of the Centaurs, such as the Edgeworth-Kuiper belt (Levison & Duncan, 1997), the Scattered Disk (Volk, K. & Malhotra, M., 2008), the inner-Oort cloud (Emel'Yanenko et al., 2005, 2007) and, more recently, the Neptune Trojans (Horner & Lykawka, 2010). Despite this uncertainty in their original source, the effect of planetary shielding on the Jupiter family comets was extensively studied by Horner & Jones (2009). Once again, their results were somewhat surprising. Rather than exclusively acting as a shield (and hence the impact rate on Earth falling as a function of "Jupiter's" mass), they found that the impact flux at Earth was again particularly low when the mass of "Jupiter" was either small or large, with a significant peak in impacts when the mass of "Jupiter" was approximately that of Saturn.

The long-period (or Oort cloud) comets (Oort, 1950) make up the final population of objects that could pose an impact threat to the Earth. These objects are sourced from a vast cloud of comets which is thought to stretch out to approximately halfway to the nearest star. The great bulk of the trillions of cometary nuclei stored in that cloud move on orbits that never bring them anywhere near the inner Solar System, effectively holding them in cold storage. However, the tidal effects of the mass of our galaxy, and the gravitational tugs and tweaks of nearby passing stars, cause some of those objects to fall on to planet crossing orbits. Many of these simply swing through the inner Solar System just once, before being ejected as a result of small gravitational nudges from the planets - Jupiter key among them. Horner, Jones & Chambers (2010) examined the effect of variations in Jupiter's mass on the frequency with which such comets were ejected from the Solar System, never to return. Clearly, any comet so ejected no longer poses any impact threat to the Earth. Perhaps unsurprisingly, the authors found that as "Jupiter's" mass increases, the efficiency with which it ejects such objects also increases, and so the resulting impact rate of long-period comets on the Earth falls away. As far as the long-period comets are concerned, then, unlike the case of the asteroids and the Jupiter-family comets, it seems that Jupiter does act as a shield, protecting us from objects that would otherwise pose a threat. However, the long-period comets constitute only a few percent of the potential Earth impactors.

The studies mentioned above are of particular interest when it comes to the question of determining a planet's potential habitability. They reveal that our long held belief that massive planets must exist outside the orbits of terrestrial worlds in order to shield them from impacts and render them habitable is, at the very least, a gross over-simplification. Indeed, in a system with no giant planets present, one obvious question is how material from any given reservoir (such as the asteroid belt or Edgeworth-Kuiper belt) can be transferred onto planet-crossing orbits.

Once exoEarths are discovered, it will clearly be critical to obtain as much information about the planetary architecture within their system as possible. Once such information is known, it would be remarkably trivial (although, of course, computationally intensive) to determine whether the planets in that system act to make it a safer or more threatening place, at least so far as impacts on planets in the classical HZ are concerned.

*Impact rates and "Late Heavy Bombardment" episodes*

Coupled to the effect of giant planets on the impact rate experienced by terrestrial worlds is the amount of material available to become potentially hazardous objects. All other things being equal, a system containing an order of magnitude more material would likely cause an order of magnitude more impacts on the planets within. Fortunately, it is possible to detect dust around other stars, utilising observations in the infra-red part of the spectrum (e.g. Harvey, Wilking & Joy, 1984; Greaves et al., 2004; Bryden et al., 2006; Greaves & Wyatt, 2010; Matthews et al., 2010). All dust within a given planetary system absorbs radiation from the host star, causing it to become heated. The warm dust then re-radiates at infra-red wavelengths, with a peak occurring at a wavelength governed by the temperature of the dust. The closer to the parent star, the warmer the dust, and so if we can detect such an excess of infra-red radiation from a given star (over that which would be expected were the star in isolation), then it is possible to make estimates of both the quantity of dust present (from the infra-red luminosity) and the location of the dust in the system (by the distribution of luminosity as a function of wavelength) (e.g. Wilner et al., 2002; Backman et al., 2009; Su et al., 2009).

There is a large difference, however, between dust and large objects. The dust which can be detected at infra-red wavelengths would be expected to have a very short lifetime around the host star before being removed by non-gravitational effects such as radiation pressure and Poynting-Robertson drag (Wyatt & Whipple, 1950). In other words, systems we observe which contain significant quantities of dust must have some source continually replenishing the dust to replace what is lost over time (e.g. Buest, 2010).

The presence of a large amount of dust around a given star is usually taken as evidence that that star is surrounded by more cometary and asteroidal objects than our own Solar System. Following that train of thought, it is often argued that exoEarths in such systems would therefore experience a significantly higher flux of impacts than that experienced here (e.g. Greaves et al., 2004; Beichman et al., 2007).

This is not necessarily the case, however. Most of the dust in the systems imaged in this way lies far beyond the orbit of the Earth. While the dust is good evidence that there is a population of objects in that region that is undergoing collisional grinding (in order to produce the observed dust), that does not necessarily mean that the system contains any efficient means to transport those bodies onto exoEarth crossing orbits. The presence of a massive disk of material at large distances from the host star might even be evidence that no giant planets exist to depopulate the disk, and so therefore the system might be more, rather than less, hospitable for life.

It is also true that a system displaying far greater concentrations of dust than our own Solar System need not necessarily contain more cometary and asteroidal bodies. In our Solar System, the asteroid belt and Edgeworth-Kuiper belts are reasonably quiescent and unstirred, having had approximately 3.8 Gyr to settle after the proposed upheaval of the Late Heavy Bombardment. During that event the great majority of unstable objects will have been removed from the belts, meaning that the belts we observe today are primarily populated by stable objects.

The Late Heavy Bombardment is a proposed heavy spike in the flux of objects passing through the inner Solar System that occurred approximately 700 Myr after the system formed. The main evidence springs from the ages of the Lunar Mare, most of which cluster tightly around the 3.8 Gyr mark. Although some people believe that this event marks simply the final tail-off in a previously higher impact regime (the end of the accretion of the Solar System, e.g. Morbidelli et al., 2001), the great majority believe that, instead, something happened to greatly increase the flux of objects passing through the inner Solar System, hugely increasing the impact flux on all the terrestrial planets, and eradicating any life present on the Earth at the time (e.g. Overbeck & Fogleman, 1989; Grieve & Pesonen, 1992; Gogarten-Boekels, Hilario & Gogarten, 1995; Wells, Armstrong & Gonzalez, 2003) . What could cause such an event? A variety of scenarios have been suggested (e.g. Wetherill, 1975; Oberbeck et al., 1977; Levison et al., 2001; Levison et al., 2008). That which has gained most recent publicity, the destabilisation of the outer Solar System as a result of planetary migration (Gomes et al., 2005), is described below, as an illustrative example.

As the giant planets formed, it is thought that they migrated, drifting through the outer Solar System as they accreted ever more material. Once the Sun cleared the solar nebula of gas and dust, a few million years after it began to form, the accretion of the giant planets was effectively over, though the terrestrial worlds likely continued to accrete from debris in the inner Solar System for at least another 100 Myr. Between the orbit of Jupiter and Mars, and outside the orbit of the furthest planet from the Sun lay reservoirs of left over material, far more massive than the asteroid belt and Edgeworth-Kuiper belt we observe today. As time passed, material was slowly removed from the outer and inner edges of these reservoirs, and passed around the outer Solar system, with the great majority eventually being ejected from the system altogether (typically by Jupiter, the most massive planet). Following Newton's third law, every time a given planet encountered such debris and exerted a force on it to change its orbit, an equal but opposite force was exerted on the planet, minutely changing its orbit in turn (indeed, using the planets to gravitationally slingshot the Voyager 2 spacecraft along its grand tour will have very slightly changed their orbits). This resulted in the planets continuing their migration, albeit far more slowly than during their initial formation.

Eventually, it is proposed, the orbits of Jupiter and Saturn migrated so far that the two planets entered a mutual mean motion resonance, and began to de-stabilise one another's orbits. This caused the outer Solar System to effectively descend into chaos. Some models even suggest scenarios in which Uranus and Neptune formed between the orbits of Jupiter and Saturn, and were ejected into the trans-planetary disk during this event. As the planets chaotically perturbed one-another's orbits, they also greatly disrupted the inner- and outer- disks, causing a great flood of asteroidal and cometary material to be thrown onto dynamically unstable orbits. Some fraction of that material was eventually thrown into the inner Solar System, causing the Late Heavy Bombardment.

Due to the chaotic nature of the event, if indeed it happened as described, it could in principle have happened earlier or later in the evolution of the Solar System than seems to be the case. During the event, it is likely that there was a greatly increased amount of dust present in our Solar System as the objects from the two disturbed reservoirs collided with one another, and everything else available.

This, then, reveals a second mechanism by which a stellar system can have a significantly enhanced amount of dust, as observed in the infra-red. If a previously stable reservoir of cometary or asteroidal material has recently been destabilised, the collision rate in that reservoir (and likely across the system as a whole) will have increased dramatically, creating much more dust than would otherwise be the case.

Should we, then, choose to target exoEarths that lie in systems with little or no observed infra-red excess? Such low levels of dust could either suggest a system containing very few potentially hazardous objects or one in which those hazardous objects that remain are all locked away in dynamically stable reservoirs. Either way, it seems that exoEarths in such systems seem a more reliable bet to have low impact rates than those in high-dust systems. While high-dust exoEarths might be equally safe (if there is no mechanism to transport the hazardous objects to intersect them), it is surely better in the first instance to look at those that are more certain to have a less hostile impact regime. That said, given that highly dusty systems might also be just as safe for a terrestrial planet as their low dust brethren, it seems that dust alone is not a particularly good criterion with which to judge the potential habitability of a system. If, however, the planetary system under consideration is known to house both a significant amount of dust, and a suite of giant planets, it would certainly be worth running dynamical integrations to attempt to model the transport of material from the dusty reservoirs to the location of the exoEarth in that system, which would allow a much stronger conclusion to be drawn on the potential habitability of the planet, and its strength as an early candidate in the search for life.

*Planetary satellites – tides and axis stabilisation*

It has often been mooted that the Moon has played a significant role in the development of life on Earth. Compared to the satellites of the other seven planets, our Moon is a surprisingly large and massive object, relative to its host planet. It is believed that the Moon formed as a result of a giant low-velocity collision between the proto-Earth and a Mars-sized object, at the end of terrestrial planet accretion (e.g. Benz, Slattery & Cameron, 1986, 1987). There is actually plenty of evidence that such giant collisions were commonplace during the final stages of planetary formation - the process is invoked to explain the anomalously high density and small size of Mercury (Benz et al., 2007), the massive impact scar that created Mars' northerly hemisphere (Andrews-Hanna et al.), the presence of Pluto's satellite system (including the giant Charon) (Stern et al. 2006), and the unusual tilt of Uranus' spin axis (Slattery, Benz & Cameron, 1992). Despite these events being common, however, they are undoubtedly random events, and there is no guarantee that any given exoEarth will have a satellite like the Moon. Could this affect its potential habitability?

One way in which our Moon has been proposed as aiding the development of life is the strong (and, when the Earth was young, rapid) tides it exerts on our oceans (e.g. Blum, 1957; Lathe, 2004; Bywater & Conde-Frieboesk, 2005). The daily ingress and egress of water on coastlines across the planet creates a vast area of land which is neither ocean nor dry land, but rather a mix of the two. It is argued that this greatly facilitated the transfer of life from the oceans to the land. It has also been suggested that this region of periodically submerged and then drying coastline could encourage the development of pre-biotic and biotic chemistry (e.g. Lathe, 2004). In the search for evidence for life, however, it is not immediately clear that having inhabited contents would make it more likely that life could be detected on a given planet - it would depend on what evidence was sought (surely oceanic life would still alter the atmosphere of the host planet, for example). However, if the Moon were not present, the Earth would still experience significant tides resulting from the pull of the Sun (the cause of the difference in height between the "spring" tides, when the Moon, Sun and Earth line up and the smaller tidal range observed when the Moon and Sun are 90 degrees apart, as seen from the Earth). As such, it is certainly possible that the processes discussed above would still occur on an exoEarth that lacked a sizeable moon. However, as discussed by e.g. Benn, 2001, the strength of the Solar tide is significantly less than the Lunar tide would have been during the emergence of life (when the Moon was far closer to the Earth than at the current time), and so the area affected by tidal activity would be proportionally less. While this would not prevent the development of life, it seems obvious that, if the role of the Moon in facilitating the development of life is accepted, then having a large satellite could definitely enhance the possibility of a given exoEarth being a suitable target in the search for life.

The other role suggested for the Moon is that it acts to stabilise the spin axis of the Earth. Over time, the tilt of the Earth's axis, relative to the plane of its orbit, rocks back and forth over a range of a degree or two - enough to cause subtle variations in climate (as discussed for the Milanković cycles, above), but not sufficient to cause catastrophic changes. In contrast, studies of Mars' obliquity have shown that the planet's axial tilt can vary wildly, sometimes even reaching (or exceeding) a tilt of 60 degrees. To put this in context, if the Earth's obliquity reached 60 degrees, the arctic circle would pass through Cairo and south of Shanghai, and to the north of Perth and Santiago in the southern hemisphere. Everything at higher latitudes than this would, at some point, experience midnight Sun and midday darkness. Beyond this, the tropics would extend to ±60° - meaning at some point in the year, the Sun would be overhead everywhere within this distance of the equator. The Earth's climate would be radically different, with incredibly harsh, sunless winters stretching most of the way to the equator, and fiery, unending summer sunshine baking everywhere within the tropics. Such conditions would clearly be far from ideal for life (although it has been suggested that the extreme excursions in Martian obliquity may lead to brief periods of time when liquid water could exist on the surface of that planet (Jakosky, Henderson & Mellon, 1995)). However, were the obliquity of such a planet constant, then the climate would at least be stable (albeit extreme) for lengthy periods of time, and it seems feasible that life could develop despite the harsh climate. It is not the extreme obliquity itself, therefore, that can be considered inimical to the development of life, but rather the speed and degree to which the obliquity changes. Clearly, a planet whose obliquity drifts chaotically between, say, 0° and 60° would present significantly greater challenge to the development of life than one, like our Earth, on which the obliquity varies over a far smaller range, or another, where the obliquity is fixed at an arbitrary angle.

It has been suggested that the main difference between the stability of the Martian and terrestrial obliquities is the result of the stabilising influence of the Moon. To study this idea in more depth, Waltham (2006) examined the idea that a large satellite can boost planetary habitability, finding that the mass of the Moon is remarkably close to the maximum for which the host planet's axial tilt would be stable. Above that mass, a large "Moon" would actually act to destabilise the spin axis of the planet, potentially making it less, rather than more, habitable. It should at this point be noted, too, that the axial tilts of Venus and Mercury are far more stable than that of Mars, despite the fact that neither planet has a massive satellite (or, indeed, any natural satellite at all!). This is likely the result of their particularly slow spin, which acts to stabilise them against major excursions in obliquity (e.g. Benn, 2001).

While the precise role played by our Moon in determining the Earth's habitability is still under debate, it seems reasonable to conclude that, should we detect exoEarths with large satellites, such a feature should be considered a plus point. Certainly, if two potential targets are equal in all other respects, aside from the presence of such a satellite, it would be reasonable to initially survey the one whose satellite most closely resembled the Earth's. Given that the formation of the Moon through impact was a stochastic event, there is no guarantee that any given exoEarth will have a large satellite (indeed, looking at the other terrestrial planets, it seems more likely that most such planets will have no major satellite). That said, should sufficient information on the dynamical state of the planetary system be available (including the presence of any satellites), it would clearly be interesting to performing calculations similar to those of Waltham (2006), to check whether the planet in question was prone to dramatic excursions in obliquity. However, given the debate which remains around this subject, we consider that the role of such satellites may well prove less significant than many of the other features discussed in this work.

## Planetary Features

*Level of hydration – planetary ocean versus desert worlds*

The presence of liquid water is often considered the key ingredient for the development of life on a planet. Indeed, the classical definition of the habitable zone is built around this assumption - the region around the star within which the temperature on an Earth-like planet would allow the presence of liquid water at some point on its surface. When we look around us on Earth, water seems to be everywhere, and so it is somewhat surprising that there is still significant debate on the original source of the Earth's water.

Theories of planetary formation suggest that the "ice line" (the closest location within the solar nebula at which water could be solid) was somewhere in the outer asteroid belt. The inner Solar System was far too hot for any water to condense out from the solar nebula, and so the terrestrial planets should have accreted from dry rocks. It may be that some water was trapped in the form of hydrated silicates, but if we assume that the terrestrial planets formed solely from the material around their current orbits, we should expect them to be almost totally devoid of volatiles. As a result of this apparent paradox, a number of different models have been proposed to detail the hydration of the terrestrial planets. These fall into two main groups.

In *endogenous* hydration scenarios, the bulk of planetary water was sourced from local materials, with some material from somewhat further away. The water was concomitantly accreted to the planet in the form of hydrated silicates, which are believed capable of storing water well within the ice-line (e.g. Drake, 2004, 2005). Although such scenarios initially seem promising from the point of view of life developing on exoEarths elsewhere, it should be noted that there is a well established correlation between the water content of meteorites that fall on the Earth and their source region in the asteroid belt. In particular, as discussed by Morbidelli et al. (2000), the family of meteorites known as Enstatite chondrites are both the driest known meteorites in the Solar System (typically just 0.05-0.1% water by mass) and the meteorites sourced from the innermost region of the asteroid belt. Such evidence suggests, then, that the amount of water delivered to the terrestrial planets during their formation from local material (such as hydrated silicates) might well have been insufficient to explain the Earth's water budget.

*Exogeneous* hydration scenarios instead suggest that the water was sourced from much further from the Sun, beyond the ice-line. These models broadly fall into two main categories. Early accretion models suggest that the water was injected into the inner Solar System (carried typically by bodies from beyond the ice-line but within the asteroid belt) whilst the planets were themselves accreting (e.g. Morbidelli et al., 2000; Petit et al., 2001). Such models often invoke the delivery through very few stochastic collisions between the planets and giant hydrated embryos. Following this logic, it is clearly feasible for an otherwise ideally placed exoEarth to either have the misfortune of receiving no such hydrating impacts, and therefore be an essentially desiccated planet. or to receive a greater number than the Earth did, and therefore be essentially an ocean world. The second category of exogeneous hydration models are often known as "Late Veneer" scenarios. These models consider late exogeneous accretion, occurring towards the end of terrestrial planet formation (e.g. Owen & Bar-nun, 1995). These models suggest that the terrestrial planets initially formed dry, and were hydrated at a later stage, possible as a direct consequence of the Late Heavy Bombardment. Such bombardment would not necessarily hydrate all the terrestrial planets equally (Horner et al., 2009) Although such models are currently somewhat out of favour, it is worth noting that the stochastic nature of the proposed Late Heavy Bombardment could mean that such a hydration event could simply not have occurred for a large number of exoEarths.

When it comes to the degree of hydration of exoEarths, the situation is complicated still further by the wide range of dynamical processes involved with the formation of systems significantly different to our own. For example, the great majority of the first exoplanets discovered were what is known as "hot Jupiters" - planets of the order of the mass of Jupiter, orbiting far closer to their parent star than Mercury does to the Sun. The discovery of these objects prompted a substantial rethink of planetary formation models (e.g. Lin, Bodenheimer & Richardson, 1996; Masset & Papaloizou, 2003; Baraffe et al., 2005). It is thought that such planets could not form on their current orbits, but would rather form much further from the parent star, and then migrate inwards, before the proto-planetary nebula is dispersed by the star. It is perfectly possible that a series of such planets could form, migrate inwards, and fall into the star before the nebula is blown away and the final wanderer frozen in place perilously close to its host. Such planetary migration does not, however, necessarily preclude the later formation of Earth-like planets on potentially habitable orbits. However, it is quite likely that the inwardly migrating Jupiters would drag with them a significant amount of hydrated material, potentially leading to the formation of hugely wet "water worlds" with oceans hundreds or even thousands of kilometers deep (e.g. Fogg & Nelson 2007). Would such a world be as suitable for the development of life as a drier, more Earth-like planet? It is often suggested that, in addition to planetary oceans, the presence of continental regions must also play an important role in the development of life. As the continents are weathered, they provide a constant source of minerals and metals that would otherwise rapidly be lost from the ocean. Without these materials, it is suggested, the development of life would be significantly stymied (e.g. Ward & Brownlee, 2000). Furthermore, if we consider that the control of planetary climate through the weathering of surface rocks plays a significant role in maintaining a suitable climate for life on Gyr timescales (as will be discussed in the next section), we remind the reader that a partially flooded planet is the only type on which such a process could act – if the planet has no oceans or no landmass, then no such weathering can occur.

So, we come to our "ideal" exoEarth. Clearly, we want to search for a planet that has liquid water on its surface, so moves within the habitable zone. However, we probably want to avoid any planet that is too dry, or too wet, and focus on those that are just right.

*Weathering, plate tectonics and the carbon cycle*

The weathering of material from the continents is thought to have played an important role in providing a wide range of important chemicals for life in the oceans that would otherwise become depleted over time. For example, calcium, which plays an important role in the functioning of cells, and is used by many creatures to produce shells, bones, and teeth, is naturally removed from the oceans over time by reaction with dissolved carbon dioxide. This produces limestone which, along with calcium and carbon sequestered in the dead bodies of sea life, settles to the ocean floor and is gradually locked away in the form of sedimentary rock. Without a source to replenish the calcium, the oceans would eventually become decalcified, which would clearly pose significant problems for life as we know it. Fortunately, whilst the calcium within the ocean is being used up, fresh deposits are introduced as continental material is slowly weathered away. The calcium deposited as limestone acts as a significant sink of carbon dioxide from the atmosphere, and can therefore play a role in modifying the planet's climate.

Although limestone plays a role in the removal and sequestration of atmospheric carbon, the most important sink of that particular greenhouse gas comes from the weathering of volcanic minerals in what is called the carbon-silicate cycle (Wallace & Hobbs (2006) b). Indeed, this process acts as an important stabiliser for the climate of the Earth. Weathering occurs more rapidly when the planet is warmer, and less so when it is cooler, primarily as a result of variations in precipitation (warm air holds more water than cold before becoming saturated). As the temperature of the planet increases, so does the rate at which the surface is weathered, which acts to remove $CO_2$ from the atmosphere at an increased rate. Over time, this causes the concentration of $CO_2$ in the atmosphere to fall, helping to slow and then eventually reverse the rise in temperatures as the greenhouse effect weakens. Similarly, should the temperature of the planet fall, the rate at which weathering removes $CO_2$ from the atmosphere also decreases. However, the rate at which fresh $CO_2$ is introduced to the atmosphere would remain roughly constant, leading to a net increase in the amount of $CO_2$ present, and, potentially, a reversal of the cooling trend.

If nothing existed to replenish the atmospheric carbon, therefore, it is conceivable that the global temperature would gradually fall until the planet itself froze over. Indeed, many authors studying the evolution of the early Earth's climate consider that our planet would have been unable to maintain sufficient atmospheric $CO_2$ to remain temperate, were that the only greenhouse gas present in the atmosphere. At the time of the Earth's formation, the Sun was only ~70% of its current luminosity, and so concentrations of greenhouse gasses must have been significantly higher than at the current day to maintain the liquid water habitat in which life developed. For this reason, it has been proposed that methane was a significant factor in maintaining the Earth's temperate climate over the first ~2 Gyr of our planet's evolution, because such weathering processes removed large amounts of $CO_2$ from the atmosphere (e.g. Lowe & Tice, 2004, and references therein).

In terms of the current day climate (ignoring anthropogenic effects), it is fortunate that plate tectonics acts to recycle sequestered carbon to the atmosphere, since any methane released to the atmosphere is rapidly destroyed. As continental plates move around the planet, fresh mountain ranges are pushed up, and the limestone from the ocean bed is exposed to the atmosphere. When limestone is weathered away, the calcium is returned to the water, and carbon dioxide released back into the atmosphere. In addition, wherever ocean floor is subducted at a plate-boundary, it melts and gives rise to volcanism (as seen, for example, along much of the Pacific "Ring of Fire"). This also acts to return material that would otherwise remain sequestered to the continental shelf, allowing the carbon cycle, and the flux of fresh material to the oceans, to continue.

Beyond its role in maintaining an appropriate mix of gases in a planetary atmosphere, it has also been suggested that plate tectonics may play an important role in maintaining convective cells whose interaction with the molten outer core of our planet plays a critical role in maintaining our planet's relatively strong (compared to the other terrestrial planets) magnetic field. The thinking goes that plate tectonics allows a planet to shed its heat significantly more quickly than the alternative "stagnant lid" tectonic setup (thought to be the case for Mars and Venus). As such, the temperature gradient within the planet is significantly greater, encouraging the transfer of that energy through convective, rather than conductive, means. Since the stagnant lid scenario slows the cooling of the mantle, it is thought it also slows the cooling of the core, which in turn prevents convection occurring on a wide enough scale to create a magnetic dynamo. It has therefore been argued that the absence of such a dynamo in Venus is the direct result of its lack of plate tectonics (e.g. Nimmo 2002).

A number of authors have suggested that the presence of water on an Earth-like planet can influence its tectonic behaviour. Indeed, were the Earth dry, they propose that it would not be able to support plate tectonics - in other words, without water, the minimum mass for a planet to house such tectonic activity would be greater than that of the Earth. The suggested lack of plate tectonics on Venus, which has a mass nearly as large as the Earth's mass, is often attributed to the planet's lack of surface water (e.g. Nimmo & McKenzie, 1998; O'Neil, Jellinek & Lenardic, 2007).

What, then, if plate tectonics on the Earth were to cease? Plate tectonics is a remarkably efficient means for the planet to lose heat, and so over time the interior of the Earth is cooling. Given enough time, the planet would cool sufficiently that plate tectonics would grind to a halt, with the Earth's tectonic state shifting to the same stagnant lid setup exhibited by Venus and Mars. This could clearly cause significant problems for the ongoing viability of the biosphere. It has been suggested that both Venus and Mars were potentially habitable in the early stages of our Solar System's evolution, although there remains some debate as to how Mars could have had liquid oceans in the early days, given the then lower luminosity of the Sun (e.g. Lunine, 1999). In both cases, however, the planets are no longer the lush oases they may once have been.

In the case of Venus, it is theorised that surface temperatures rose sufficiently that they passed a certain "critical point" (potentially just ~50°C, or so), above which large amounts of the surface water evaporated. Water is itself a strong greenhouse gas, and having a very humid atmosphere would have helped Venus continue to warm. Additionally, as the water was carried high into the planet's atmosphere, it was dissociated by Solar UV radiation (which is more intense at Venus than the Earth, due to its proximity to the Sun, and was also being radiated in significantly greater quantities at that time, as discussed earlier). This process was undoubtedly aided by the structure of the Venusian atmosphere. On Earth, our atmosphere has a strong temperature inversion above the troposphere, which acts to keep the great bulk of our planet's atmospheric water contained well below the altitude needed for photo-dissociation. It is quite plausible that Venus did not have such an inversion, which would have resulted in the gradual bleeding of the planet's oceans away into space. As the temperature rose, and the planet dried, any plate tectonics on the surface would have ceased (Ward & Brownlee, 2000). This, in turn, would slow the release of any water trapped in the planet's mantle (through volcanic activity), and may well have led to the collapse of any strong magnetic dynamo that the planet may once have possessed. At the current epoch, Venus has little or no measurable magnetic field ($< 10^{-8}$ T, according to Stevenson, 2003, at least a factor of a thousand weaker than that at the Earth).

In the case of Mars, a number of routes have been proposed to explain the loss of the planet's initial thick atmosphere. There is a substantial amount of evidence that liquid water once flowed upon the planet, which shows that the conditions on the surface were once temperate. However, most authors agree that the "wet Mars" phase lasted less than a billion years. There is some evidence that flowing water has existed on a temporary basis since that time (usually related to impacts, volcanism, or avalanches removing the pressure on subsurface ice deposits). However, with such a thin atmosphere and lack of an effective greenhouse, conditions are such that any surface water today would rapidly freeze or evaporate. The removal of the martian atmosphere by impacts, aided by Mars's low gravitational field, could easily have contributed to the rapid transition from "wet" to "dry", lowering the atmospheric pressure and removing significant quantities of the planet's water at the same time. Equally, it seems likely that inorganic processes would have removed large amounts of $CO_2$ from the martian atmosphere, trapping it in clays beneath the planet's postulated oceans and lakes. This would clearly hasten the cool-down of the planet. Being significantly smaller than the Earth or Venus, Mars has a higher surface area to volume ratio, which means it loses heat significantly more quickly than the larger two planets (in fact, this is the same reason that small animals such as mice suffer more severely from cold weather in winter than humans do!). Such rapid cooling would hasten the demise of any martian plate tectonics, once again removing a mechanism to facilitate the recycling of greenhouse gases. In addition, because Mars is less massive than the Earth, its gravitational field is proportionally less, resulting in a lower escape velocity. Because of this, any $N_2$ which was dissociated by the effect of solar UV radiation would then bleed away into space, significantly reducing the planet's atmospheric pressure. As all these factors led to the planet cooling, volatiles in the atmosphere would have begun to freeze out, sequestering vast quantities of $CO_2$ and water as ices, trapped at the polar caps and in the subsurface. Finally, as we discuss in the next subsection, we note that, if the primordial martian magnetic field was significantly weaker than that of the Earth, or shut down particularly early in the planet's evolution, this too would act to speed the loss of the planet's atmosphere.

*Magnetic field*

All stars, even those that are fairly quiescent (such as our Sun), continually expel a prodigious amount of material (in the form of their stellar wind and more violent coronal mass ejections), primarily in the form of charged particles. Unimpeded, this material would directly interact with the atmospheres of the terrestrial planets, resulting in the gradual but unceasing erosion of the atmosphere. Fortunately, the Earth has a particularly strong magnetic dynamo, which acts to shield the atmosphere from the worst vagaries of the Sun's influence. Still, some of the solar charged particles make it through the field, and then impact upon the atmosphere to produce the beautiful Aurora Borealis and Aurora Australis. Without the strong magnetic field, the flux of such material would be significantly higher.

The role of a magnetic field in preventing atmospheric loss is particularly important during the early part of a host star's life. As time goes on, the strength of a stellar wind decreases, and the efficiency with which it could remove a planet's atmosphere diminishes. In the case of Mars, spacecraft observations suggest that the planet's magnetic dynamo shut down around 4 Gyr ago. At that time, it is thought that Mars had sufficient atmosphere to maintain liquid water on the surface, the source of many features we see today. Without the protection of a magnetic field, however, the combined effects of the solar wind and the freeze-out of $CO_2$ from the Martian atmosphere have played a significant role in withering away its atmosphere until we are left with the current atmosphere of just ~7 millibar (hereafter mb) average surface pressure (e.g. Dehant et al., 2007, who discuss the role of planetary magnetic dynamos on the protection of the early atmospheres of Earth and Mars in some depth).

In contrast, however, Venus has a far thicker atmosphere than the Earth, despite having a far weaker magnetic field, and lying significantly closer to the Sun. It therefore experiences a significantly stronger Solar wind than either the Earth or Mars. Taken in isolation, these facts would appear to suggest that the planet should have lost the great bulk of its atmosphere, and have been left almost an airless husk as a result of the early Sun's exuberant behaviour, since the erosive effect of the Sun's activity would have been far, far higher during the star's youth than at the current time.

However, it seems highly likely that plate tectonics was active on the youthful Venus, and that it may well have lasted for a significant fraction of its life (although it is not in operation today). Assuming that Venus held on to its water, and experienced plate tectonics, for the first few hundred million years (or even the first billion years) of its early life), it is reasonable to assume that it would have also maintained a strong enough magnetic field to have escaped the worst vagaries of Solar erosion. Once the oceans boiled, and the planet's water was lost, efficient removal of $CO_2$ by the weathering of volcanic deposits would have ceased, while the outgassing of $CO_2$ from the interior of the planet would have continued, resulting in a gradual growth of the planet's atmosphere. While the Sun is clearly acting to slowly remove Venus atmosphere, its middle-aged activity is low enough that the rate of atmospheric loss is small, which has allowed Venus to maintain its massive atmosphere to the current epoch. As stated in by Lundin, Lammer & Ribas (2007), "Mars and Venus represent two extremes of the consequence of un-shielding". Indeed, those authors find that their model of solar forcing is "sufficiently effective to remove some 40 bar of water from Mars and at least 50 bar of water from Venus". For a more detailed discussion of the effect of solar forcing on the atmospheres of Venus, Earth, and Mars, we direct the interested reader to their work.

In addition to slowly stripping the atmosphere of the planet away, the increased flux of Stellar material impinging on planetary atmospheres could hinder the development of advanced life in a number of other ways. Scientists studying some of the more intense solar storms of the last few decades have noted that the most intense can cause a small depletion in the Earth's ozone layer, as solar protons dissociate some of the molecules in the upper atmosphere, freeing significant amounts of material that can react with the ozone, reducing the ozone layer (e.g. Jackman et al., 2001). Fortunately, thanks to the presence of the Earth's magnetic field, only the most energetic protons make it through to the atmosphere, even in the most intense storms. Without the field, it seems likely that enough solar material would make it into the upper atmosphere to render the development of a protective ozone layer almost impossible.

Of the terrestrial planets, the Earth has by far the strongest magnetic field (e.g. Stephenson 2003, Table 1), and as described above, this is viewed as having played a key role in the creation of a habitable environment for life to develop. As described earlier, the magnetic field of the Earth could well be strongly linked to tectonic processes, which help to maintain sufficient convection within the mantle and outer core to create our planetary dynamo. Given that it is possible that, were our Earth a dry planet, rather than a wet one, plate tectonics would not be possible, it seems likely that, for planets of around the mass of the Earth, the presence of water could play a far more critical role in determining habitability than simply providing a solvent in which life can develop. Although more massive exoEarths could maintain such tectonic activity even if they were significantly drier than the Earth, this adds further weight to the idea that the first truly Earth-mass planets we search for life must have a significant water budget. For a more detailed in-depth discussion about the effect of planetary magnetic fields on habitability, we direct the interested reader to section 4.3 of the excellent review by Lammer et al. (2009).

*Atmospheric pressure, structure, and orbital distance*

The atmosphere of any exoEarth will play a key role in determining whether the planet is habitable. A few of the reasons for this have already been mentioned, but key among them is the role of an atmosphere in ensuring that the planet's surface is capable of sustaining liquid water. There has to be sufficient atmospheric pressure that liquid water is a possibility, as a start. Below a pressure of 6.10 mb, liquid water cannot exist. Any lower, and ice passes straight to the vapour phase, without ever being liquid. Interestingly, 6.10 mb is very close to the mean atmosphere pressure on the surface of Mars. Above a certain martian altitude, liquid water can never be stable, regardless of the local temperature. So, clearly, a planet must have a significant atmosphere in order to house liquid water. As the atmospheric pressure on a planet increases, so does the upper temperature at which water remains liquid. On Earth, the mean atmospheric pressure at sea level is 1013 mb, and pure water is liquid from 0 °C up to its boiling point at this pressure of 100 °C.

Beyond providing enough pressure to allow liquid water to be present, the atmosphere also plays a key role in maintaining an optimal temperature for that water. Too cold, and the water freezes out, too warm, and it boils away (and is eventually potentially lost as photo-dissociation in the planet's upper atmosphere breaks it to its component hydrogen and oxygen). This balance between too hot and too cold is not so simple as it might seem. Remember that the luminosity of a star gradually increases throughout its main sequence lifetime. Our Sun was just 70% of its current luminosity when it entered the main sequence, and it is clear that with the Earth's current atmospheric makeup, it would have been far too cold to host liquid oceans - a snowball Earth. Fortunately, the Earth's early atmosphere was significantly different to that we see today, with huge quantities of greenhouse gases (primarily $CO_2$ and $CH_4$) raising the temperature enough to support our oceans. Indeed, it is believed that the temperature of the early Earth was significantly higher than it is today.

Just as the present Earth's atmosphere would have led to the early Earth freezing over, if the modern Earth had retained the early Earth's atmosphere, the greenhouse effect would have long ago boiled the planet's oceans, and left us a dry, overheated husk (likely much like Venus). So the atmosphere, then, must evolve in such a way to maintain liquid water on the planet's surface over Gyr timescales. At this point, the inter-relationship between the atmosphere, weathering, and the effects of life itself become incredibly complicated - but it is certainly fair to say that if we were searching for the "best" exoEarth to target, it would have to have an atmosphere today that would support the presence of liquid water over a significant fraction of the planet's surface.

The atmosphere of the Earth has played an additional important role in the maintenance of the oceans. Uniquely among the terrestrial planets, our atmosphere has a significant temperature inversion above the troposphere. This unusual vertical temperature structure proves remarkably efficient at preventing any significant quantities of water diffusing high enough in the atmosphere to become dissociated. Without that inversion, our planet would rapidly shed hydrogen to space, as the water is broken down, and the oceans would slowly be lost.

At the other end of the scale, there comes the more speculative question of how much atmosphere is too much? As the atmospheric pressure on a planet increases, so does the range of temperatures at which water remains liquid. Therefore, one could imagine massive, hot exoEarths being able to house significant oceans. Would such a thick atmosphere prove prohibitive to the development of life? Could it be that such an atmosphere would make it harder for any life on that planet to be detected?

On the other hand, increasing the thickness of a planet's atmosphere (which could perhaps be related to an increased mass of the planet itself - potentially super-Earths would have super-atmospheres) would allow the planet to retain more of the heat it receives from the Sun, potentially increasing the radial extent of the classical habitable zone, and allowing planets that would otherwise be excluded to be considered potentially habitable. As an example in our own Solar System - if Mars, which has a mass about a tenth that of the Earth, were at least a few times more massive than it is, then it is likely it would have retained a significantly thicker atmosphere, and also been able to maintain plate tectonics (assuming Mars was wet!). In such a scenario, it seems reasonable to think that such a "Mars" would be habitable at the current epoch, despite its position on the very outer edge of the present day classical habitable zone.

## Conclusion

If the history of astronomy teaches us anything, it tells us that as technology improves, new populations of objects will be discovered. Although the search for the first in a given population is arduous, and takes many years, once one is found, many others follow soon after. In the last couple of decades, such rapid growth after initial discovery has been observed time and time again - witness the rate at which exoplanets have been found since the discovery of 51 Pegasi, and the rapid growth in the number of trans-Neptunian objects and Centaurs known within our own Solar System.

In the coming years, the first truly Earth-like exoplanets will be discovered - the exoEarths. Just as with every other population of objects discovered by astronomers, it is certain that where one discovery leads, many others will quickly follow. Once these planets are found, the scramble to search for the first concrete evidence for life beyond the Solar System will begin in earnest. Such a search will be incredibly challenging, with observations pushing telescopes and observers far beyond anything seen to date. In addition, however, those observers will want to be extra-certain of their results before publishing, giving the incredible weight of any possible discovery. For those reasons, it is certain that it will be almost impossible to simultaneously survey the entire catalogue of exoEarths in sufficient detail to claim a certain detection of life. Rather, the few most promising candidates will be hand-picked for the first suite of observations. There are a wide range of factors which could render an otherwise hospitable planet less so, and in deciding which planets to concentrate the search on, it is important to take into account as many as possible.

In this review, we have illustrated many of the various effects which can influence the habitability of an exoEarth. In the coming years, it is imperative that scientists from all fields within astrobiology come together to prepare a template for "optimal habitability", to help determine which exoEarth should be the first target for an intensive search for life. Some of the processes long held to be vital for the presence of life might prove, under further research, to be less important than previously thought, while others might be considered show-stopping. Nevertheless, we potentially have less than a decade to prepare the tools and our blueprint for habitability, to ensure that we are ready to focus our efforts once the first exoEarths are found.